\documentclass[conference]{IEEEtran}

\pdfoutput=1
\usepackage{graphicx}
\usepackage{epsfig}
\usepackage{amsmath}
\usepackage{amsfonts}
\usepackage{enumerate}

\usepackage{graphicx}

\begin{document}

%
\title{Game Theoretic Network Coding-aided MAC for Data Dissemination towards Energy Efficiency}

\author{\IEEEauthorblockN{Angelos Antonopoulos and Christos Verikoukis}
\IEEEauthorblockA{Telecommunications Technological Centre of Catalonia (CTTC)\\
Castelldefels, Barcelona, Spain\\
Email: \{aantonopoulos, cveri\}@cttc.es}\\
}

\maketitle              

\begin{abstract}
In this paper we propose game theoretic Medium Access Control (MAC) strategies for data dissemination scenarios. In particular, we use energy-based utility functions that inherently imply power-awareness, while we consider network coding techniques to eliminate the necessity of exchanging acknowledgement control packets. Simulation results show that our proposed strategies enhance the energy efficiency of the system and reduce the dissemination completion time compared to an optimized standard protocol.
\end{abstract}



\begin{IEEEkeywords}
Medium Access Control (MAC); Nash Equilibrium; Network Coding; Energy Efficiency.
\end{IEEEkeywords}

\IEEEpeerreviewmaketitle

\section{Introduction}
\label{sec:intro}
\IEEEPARstart{D}{ata} dissemination concept attracts increasing attention, since the spreading of digital information becomes of paramount importance. Such information could be multimedia files or continuous streaming segmented into packets to be shared properly among the interested nodes. The problem of data dissemination is harder and more complicated in Mobile Ad Hoc Networks (MANETs) due to the lack of any infrastructure and the instability of wireless links. However, network coding \cite{IEEEhowto:nc1} has been recently introduced in order to provide the communication with robustness, diversity and enhanced Quality of Service (QoS).

Network coding has been proven to achieve the multicast capacity \cite{IEEEhowto:nc2}, thus being potentially a beneficial application for data dissemination. Furthermore, since data collection is an equivalent situation to the coupon collector's problem \cite{IEEEhowto:coupon}, network coding can significantly simplify the complexity of the solution \cite{IEEEhowto:deb}. Specifically, transmitting linear combinations of the packets instead of just forwarding the information flows eliminates the necessity of exchanging acknowledgements. Hence, it is sufficient for a network node to receive enough independent linear packet combinations in order to decode the entire data set.

Recently, it has been shown that giving priority to the node with the most information does not speed up the dissemination process. Specifically, Lucani et al. \cite{IEEEhowto:lucani} demonstrated that the completion time is minimized by selecting the node with the greatest impact\footnote[1]{Impact is defined as the number of innovative packets that are delivered to the sink nodes in one transmission.} in the network to transmit in each slot. However, in their paper it is neither defined any realistic MAC protocol nor specified which node transmits when more than one node has the greatest impact in the network.

The work of \cite{IEEEhowto:lucani} has motivated us to design medium access strategies for data dissemination scenarios. Let us recall that the upper goal of data dissemination is to share a number of packets among a set of sink nodes. Specifically, the goal of all nodes is twofold: i) to complete the dissemination in a reasonable time and ii) to maximize their lifetime. Therefore, during the dissemination process, the source nodes have to balance a trade-off between transmitting data and saving energy. This fact, along with the selfish nature of the nodes, has inspired us to consider game theory \cite{IEEEhowto:gintis} as an appropriate tool in order to provide this problem with a reliable solution.

In this paper, we present energy efficient game theoretic medium access strategies for data dissemination. We introduce a Dissemination Access Game (DAG) to identify a state of balance (\emph{equilibrium}) of the network nodes between saving energy and proceeding the dissemination procedure. The contribution of our proposed scheme lies on the following:

\begin{enumerate}
\item We use energy-based game theoretic techniques to improve the energy efficiency of the system without compromising the offered QoS.
\item We propose a general access scheme that can be applied in several wireless Standards.
\end{enumerate}

The rest of the paper is organized as follows. In Section \ref{sec:dag} we introduce the two versions of our proposed Dissemination Access Game (simple access and delay-bounded access strategy), along with the game theoretic framework of the problem. The performance evaluation of our protocols is provided in Section \ref{sec:performance}, while Section \ref{sec:conclusion} concludes the paper.

\section{Proposed Medium Access Game for Data Dissemination}
\label{sec:dag}

\subsection{System Model}

\begin{figure}[htb]
\centering
\includegraphics[width=1\columnwidth]{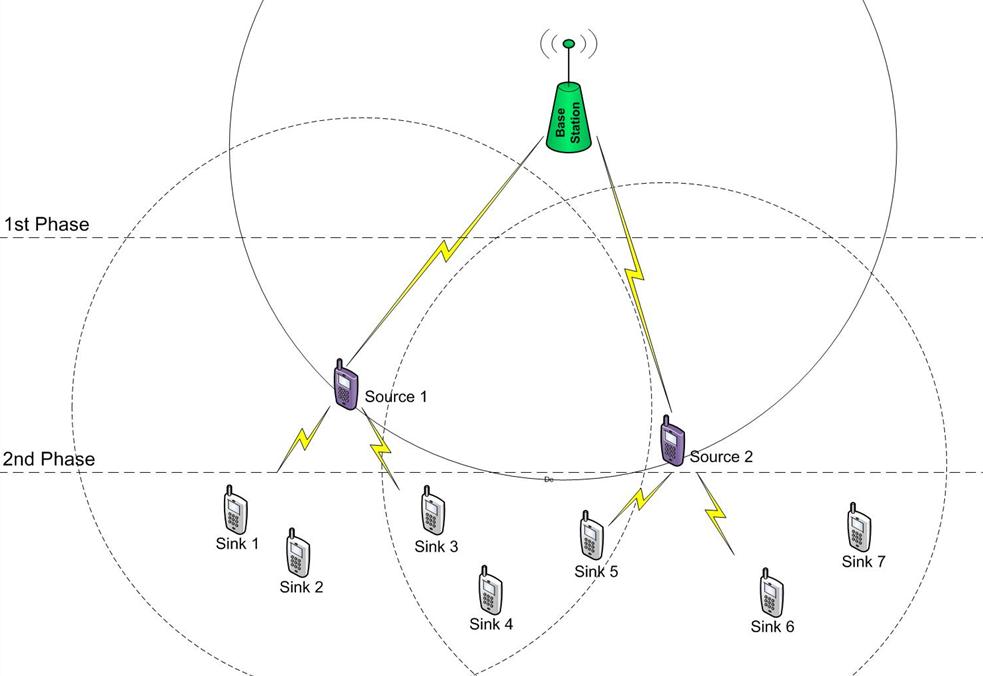}
\caption{System Model}\label{f10}
\end{figure}

We consider a network topology with a base station that holds the total amount of information and a set of nodes that desire the disseminating data. The dissemination takes place in two phases: i) in the first phase the base station broadcasts the information to the nodes that are placed inside its transmission range, and ii) in the second phase the nodes that have already received the information (so called source nodes) are enabled to forward the data to the rest interested nodes (so called sink nodes). We focus on the second phase of the problem, where we consider that there are two sources that have already obtained the total amount of information and a set of $L$ sink nodes that are interested in the disseminating data (Fig. \ref{f10}). Furthermore, the transmission ranges of the sources are partially overlapped, while both sources affect the same number of sink nodes, thus having the same impact in the network. We further assume a slotted system, where the node with the maximum impact on the network gets the access to the channel in order to transmit in each slot. Apparently, the existence of more than one source in the network generates conflicting situations that need to be solved by specific medium access control mechanisms.

Regarding the data transmissions, random linear network coding techniques are adopted to facilitate the data dissemination. In particular, the nodes transmit linear combinations of the packets, thus eliminating the need of control packets (i.e. acknowledgements). However, an extra overhead is added to the packets, since the network coding header includes information which is necessary for the decoding of the packets, such as the coding vector, the generation size and the generation identifier.

\subsection{Game Model}
\label{sec:game}

In order to focus on the energy aspect of the problem, we choose the utility function such that to quantify the lifetime of the nodes. Defining $\mathcal{E}_{TOTAL}$ as the total energy amount available to each node and $\mathbf{E}[\mathcal{E}]$ as the average amount of energy that is consumed by the wireless interface in each slot due to transmissions, the utility function of client $i$ is given by $U_i=\frac{\mathcal{E}_{TOTAL}}{\mathbf{E}[\mathcal{E}_{i}]}$. An ideal scenario would be a round robin scheduling among the source nodes coordinated by a central controller or explicit information exchange between the nodes. However, this would induce more interference and require extra energy consumption, while it is applicable only in infrastructure networks.

Since the global goal is the completion of the dissemination, all nodes have profit of transmitting data. However, the sender's role implies energy wasting, hence no client would take up this role, unless particular incentives would be provided. On the other hand, if no one transmits, the nodes will end up with spending all their energy in waiting. To analyze this conflicting situation, we model the access scenario as a \emph{static non-cooperative game with complete information}, where each node maximizes its own utility.

In order to be compatible with the notations of game theory, let us start by generally denoting the set of the players\footnote[3]{Note that the terms "player", "node", "client" will be used interchangeably in this paper.} as $N=\{1,2\}$. The action set includes two feasible actions, i.e. $A=\{transmit(T),wait(W)\}$. However, in order to deal with mixed strategies, we define as actions the transmit probabilities of the players, i.e. $s_i=\alpha_i(T),\forall i\in N$. Therefore, the extended strategy space of player $i$ is given by $S_i=\{s_i|0\leq s_i\leq 1\}$. The strategy combination is denoted as $\textbf{s}=(s_1,s_2)\in S$, where $S$ is the Cartesian product of the two players' strategy spaces. Furthermore, we define as:

\begin{equation}
\label{eq:s-i}
 	\textbf{s}_{-i}=(s_1,...,s_{i-1},s_{i+1},....,s_n)\in S_{-i}
\end{equation}
the strategy combination of all players except $i$. Moreover, we extend the definition of the utility function and we use $U_i(\textbf{s})$ to denote the utility of player $i$ when the strategy combination is $\textbf{s}$. Finally, since we deal with mixed strategies, a strategy combination $\textbf{s}^{*}$ is said to achieve the Nash Equilibrium Point ($NEP$) \cite{IEEEhowto:nash} when:

\begin{equation}
\label{eq:nash}
 U_i(\textbf{s}^{*})\geq U_i(\textbf{s}_{-i}^{*},s_i),\forall s_i\in S_i, i\in N
\end{equation}

\subsection{Dissemination Access Game}
\label{sec:dag1}
In the following subsections we introduce two versions of our proposed Dissemination Access Game: i) a simple access strategy, where the players estimate the $NEP$ according to energy-based utility functions and ii) a delay-bounded access strategy, where the players act in a way similar to the simple access scheme, while a central controller is occasionally used to reduce for the completion time of the dissemination.
\subsubsection{Simple Access Scheme}
\label{sec:scheme1}
\begin{figure}[htb]
\centering
\includegraphics[width=1\columnwidth]{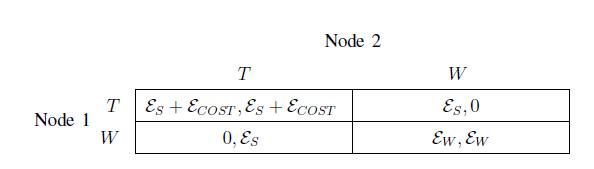}
\caption{The cost matrix of the proposed game.}\label{game}
\end{figure}

The straight-forward representation of the game in its strategic form is presented in Fig. \ref{game}. The set of the feasible pure actions is $A=\{transmit(T),wait(W)\}$, while the contents of the table are energy costs that each node wishes to minimize.


Three different cases derive from Fig. \ref{game}:
\begin{enumerate}
\item \textbf{Both nodes transmit:} The nodes waste energy for the transmissions ($\mathcal{E}_{S}$), while the collision of the packets adds an extra cost, $\mathcal{E}_{COST}$, since the dissemination does not proceed.
\item \textbf{One node transmits - One node waits:} The transmitting node wastes energy for the transmission ($\mathcal{E}_{S}$), while the backoff node has a zero-consumption since the data dissemination proceeds normally.
\item \textbf{Both nodes wait:} The nodes do not spend energy on transmissions but they do have an extra cost,  $\mathcal{E}_{W}$, since they expend energy while the dissemination does not proceed.
\end{enumerate}

Since there is no efficient equilibrium in pure strategies, each node selects a transmit probability, $s_i$, independently of the others. Therefore, the expected energy wasted for the two nodes is given by\footnote[3]{We use the notation $\bar{s}_i$ to denote the complementary probability of $s_i$, i.e. $\bar{s}_i=1-s_i$.}:
\footnotesize
\begin{equation}
\label{eq:e1}
 	\mathbf{E}[\mathcal{E}_{1}]=s_1 \cdot \bar{s}_2\cdot \mathcal{E}_{S}+ s_1 \cdot s_2 \cdot (\mathcal{E}_{S}+\mathcal{E}_{COST})+\bar{s}_1 \cdot \bar{s}_2\cdot\mathcal{E}_{W}
\end{equation}
\begin{equation}
\label{eq:e2}
    \mathbf{E}[\mathcal{E}_{2}]=s_2 \cdot \bar{s}_1\cdot \mathcal{E}_{S}+ s_2 \cdot s_1 \cdot (\mathcal{E}_{S}+\mathcal{E}_{COST})+\bar{s}_2 \cdot \bar{s}_1\cdot\mathcal{E}_{W}
\end{equation}
\normalsize
where the symmetry of the equations motivates us to search for symmetrical strategies.

Furthermore, the indifference principle defines that the mixed strategy $NEP$ of a game can be simply computed by making each player indifferent among his strategy choices and, hence, we have to estimate the roots of the partial derivative of the expected cost of player 1 with respect to $s_2$, i.e. $\frac{\partial \mathbf{E}[\mathcal{E}_1]}{\partial s_2}=0$.

For simplicity reasons and without loss of generality, let us assume that $\mathcal{E}_W=a \cdot \mathcal{E}_S$ and $\mathcal{E}_{COST}=b \cdot\mathcal{E}_S$. Consequently:

\begin{equation}
\label{eq:ne}
\mathcal{E}_{S}\cdot(a\cdot(s_1-1)+b\cdot s_1)=0\Longrightarrow s_1=\frac{a}{a+b}
\end{equation}

\begin{figure}[htb]
\centering
\includegraphics[scale=0.5]{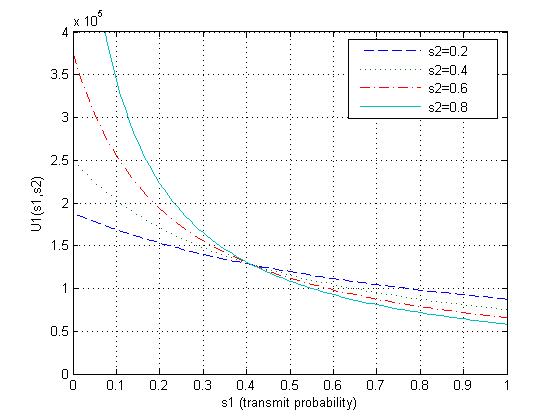}
\caption{Player 1's Utility vs. $s_1$}\label{f2}
\end{figure}

Given that the power level of the idle state in IEEE 802.11 is the 70\% of the transmission state \cite{IEEEhowto:ebert}, we set $a=0.7$, while we assume that $b=1$. Fig. \ref{f2} shows a plot of client 1's utility for various values of $s_2$, using numerical values: $\mathcal{E}_{TOTAL}=100J$, $\mathcal{E}_{S}=9.5\cdot 10^{-4}J$, $\mathcal{E}_{COST}=9.5\cdot 10^{-4}J$ and $\mathcal{E}_{W}=6.7\cdot 10^{-4}J$.

In this figure we can do the following observations:
\begin{itemize}
\item The maximum utility of node 1 increases with $s_2$.
\item For large values of $s_2$, the optimal strategy for node 1 is to wait.
\item If node 1 transmits with $s_1\simeq0.412$, its utility is independent of node's 2 strategy, $s_2$. Since the game is symmetric, the same holds for player 2. This value verifies our analysis (Eq. (\ref{eq:ne})) for $a=0.7$ and $b=1$.
\end{itemize}

 Hence, considering the symmetry of the game, we conclude that the strategy $\textbf{s}=(s_1\simeq0.412,s_2\simeq0.412)$ is the $NEP$ of our game, since any other strategy would violate the indifference principle.

\subsubsection{Delay-bounded Access Scheme}
\label{sec:scheme2}
Applying the proposed simple access scheme to the network, it is possible to have unsuccessful/empty slots in the network either due to collisions or idle slots when the nodes mutually transmit or wait, respectively. Hence, in order to bound the time that is needed to complete the data dissemination, we adopt the use of a central controller that defines which node is going to transmit, when we have two consecutive slots without any successful transmission. However, it is worth noticing that the controller only senses the channel and intervenes occasionally by polling one station, hence eliminating the necessity of control packets and extra overhead.

In this case, we have the following contingent before having a successful transmission of a packet:
\begin{enumerate}[i)]
\item $1^{st}$ slot: either successful or unsuccessful transmission.
\item $2^{nd}$ slot: unsuccessful transmission in the first slot followed by either successful or unsuccessful transmission.
\item $3^{rd}$ slot: unsuccessful transmissions in the first two slots and the central controller defines which node is going to transmit.
\end{enumerate}

Therefore, the expected energy cost for the node 1 is given by:
\footnotesize
\begin{equation}
\label{eq:E-1}
\begin{aligned}
&\mathbf{E}[\mathcal{E}'_{1}]=[s_1 \cdot \bar{s}_2\cdot \mathcal{E}_{S}+ s_1 \cdot s_2 \cdot (\mathcal{E}_{S}+\mathcal{E}_{COST})+\bar{s}_1 \cdot \bar{s}_2\cdot\mathcal{E}_{W}]+\\
&+[\bar{s}_1 \cdot \bar{s}_2^2\cdot s_1\cdot \mathcal{E}_{S}+\bar{s}_1 \cdot \bar{s}_2\cdot s_1 \cdot s_2 \cdot (\mathcal{E}_{S}+\mathcal{E}_{COST})+\bar{s}_1^2 \cdot \bar{s}_2^2\cdot \mathcal{E}_{W}\\
&+ s_1^2 \cdot s_2 \cdot \bar{s}_2\cdot\mathcal{E}_{S}+s_1^2 \cdot s_2^2 \cdot (\mathcal{E}_{S}+\mathcal{E}_{COST})+s_1 \cdot s_2 \cdot\bar{s}_1 \cdot \bar{s}_2\cdot\mathcal{E}_{W}]+ \\
&[\bar{s}_1^2 \cdot \bar{s}_2^2\cdot p_{poll} \cdot \mathcal{E}_{S}+s_1^2 \cdot s_2^2\cdot p_{poll} \cdot \mathcal{E}_{S}+2\cdot s_1 \cdot s_2 \cdot\bar{s}_1 \cdot \bar{s}_2\cdot p_{poll} \cdot \mathcal{E}_{S}]\\
\end {aligned}
\end{equation}
\normalsize
where the three terms in brackets represent the expected value of energy consumption with regard to the three aforementioned cases, respectively. In order to search for equilibrium strategies, let us assume again that $\mathcal{E}_W=a \cdot \mathcal{E}_S$ and $\mathcal{E}_{COST}=b \cdot\mathcal{E}_S$, while the probability $p_{poll}$ is constant, due to the fairness of the scheduler.
Differentiating the utility function $U'_1=\frac{\mathcal{E}_{TOTAL}}{\mathbf{E}[\mathcal{E}'_1]}$ with respect to $s_1$ gives:

\begin{equation}
\label{eq:diff}
\frac{\partial U'_1}{\partial s_1}=-\frac{\mathcal{E}_{TOTAL}}{(\mathbf{E}[\mathcal{E}'_1])^2}\cdot \frac{\partial (\mathbf{E}[\mathcal{E}'_1])}{\partial s_1}
\end{equation}

The best response of $s_1$ to the strategy $s_2$ is given by setting $\frac{\partial U'_1}{\partial s_1}=0$ or equivalently (from Eq. (\ref{eq:diff})) $\frac{\partial (\mathbf{E}[\mathcal{E}'_1])}{\partial s_1}=0$.
Therefore

\footnotesize
\begin{equation}
\label{eq:br}
\begin{aligned}
&\bar{s}_2\cdot b+2\cdot s_2\cdot b-\bar{s}_2\cdot a \cdot b-\bar{s}_2^2 \cdot s_1 \cdot b-\bar{s}_1\cdot \bar{s}_2^2 \cdot b+\\
&+2\cdot \bar{s}_1\cdot \bar{s}_2\cdot s_2 \cdot b-2\cdot\bar{s}_1\cdot \bar{s}_2^2 \cdot a \cdot b+6 \cdot s_1\cdot s_2^2\cdot b+\\
&+s_2 \cdot \bar{s}_1\cdot \bar{s}_2 \cdot a \cdot b-s_1 \cdot s_2 \cdot \bar{s}_2\cdot a\cdot b=0\\
\end{aligned}
\end{equation}
\normalsize

\begin{figure}[htb]
\centering
\includegraphics[scale=0.5]{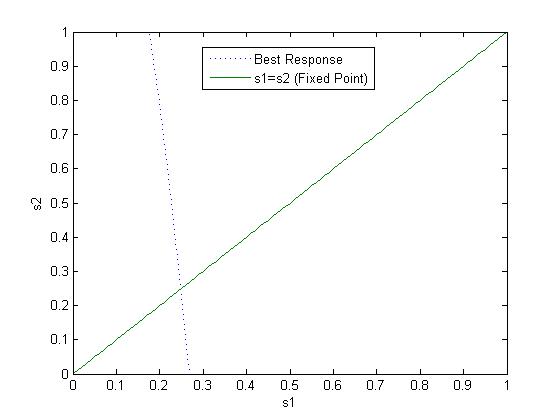}
\caption{$s_1$ vs $s_2$ and Fixed point solution}\label{f3}
\end{figure}

Figure \ref{f3} shows a plot of Eq. (\ref{eq:br}) when $a=0.7$ and $b=1$. It can be observed that there is only one fixed-point solution in the strategy space $(s_1,s_2 \in [0,1])$, i.e. $s_1^*=s_2^*\simeq0.25$, which is the unique equilibrium of the game. In order to justify this, we use the definition of the $NEP$. Specifically, since the $NEP$ is defined as the mutual best response, we conclude that any fixed point in mixed strategies constitutes a $NEP$ \cite{IEEEhowto:mackenzie}.

\section{Performance Evaluation}
\label{sec:performance}
We have developed a time-driven C++ simulator that implements the rules of the proposed DAG schemes and we have carried out Monte Carlo simulations to evaluate the performance of the protocols. In this section, we present the simulation set up along with the results of our experiments.

\subsection{Simulation Scenario}
\begin{figure}[htb]
\centering
\includegraphics[width=1\columnwidth]{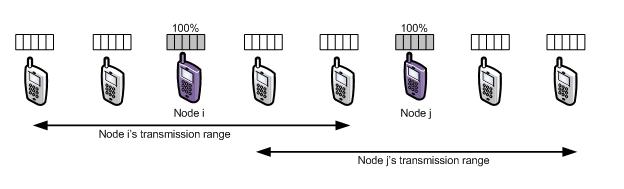}
\caption{Simulation Scenario}\label{f5}
\end{figure}

The network under simulation consists of eight nodes in total, where two of them have already obtained the total amount of information broadcasted by the base station, while the rest six are sink nodes (Fig. \ref{f5}). During the dissemination, the two source nodes have the same impact on the network, since they affect the same number of sink nodes (four sink nodes in the range of each source). Additionally, the nodes are capable of performing network coding techniques to their buffer packets before forwarding them. Hence, the necessity of exchanging acknowledgements is eliminated, since it is sufficient for each node to receive a specific number of independent linear combined packets. In our simulations we assume that the nodes need to obtain 192 linearly independent packet combinations to extract the total information. The coding of the packets is performed over a finite Galois Field - $GF(2^8)$, since it has been proven to be sufficient for linear independence among the packets. The specific field implies that the number of the encoding packets represents the number of the bytes in the encoding vector. If we use one generation of $192$ packets, the extra overhead in each packet will be $192$ bytes, which is huge especially for small size payloads. Therefore, we have chosen to create $16$ generations of $12$ packets each, which results in a network coding header of $13$ bytes in total ($12$ bytes for the encoding vector, $4$ bits for the generation size and $4$ bits for the generation identifier).

The time slot has been selected equal to 20$\mu$sec according to the IEEE 802.11g physical layer \cite{IEEEhowto:80211g}, while we consider two different transmission rates with regard to the Signal-to-Interference-plus-Noise-Ratio (SINR) values: i) 54 Mb/s for \emph{high SINR} conditions and ii) 24 Mb/s for \emph{low SINR} conditions. Regarding the power, Ebert et al. \cite{IEEEhowto:ebert} have measured the power consumption of a wireless interface during the transmission and reception phase. Based on their work, we have chosen the following power levels for our scenarios: $P_T=1900mW$ (power consumed in \emph{Transmission Mode}), $P_R=P_I=1340mW$ (power consumed in \emph{Reception} and \emph{Idle Mode}, respectively). The value of $P_T$ has been selected as an average value of transmission consumed power, since it varies according to the Radio Frequency (RF) power level.

In our experiments, we compare the proposed strategies with a IEEE 802.11-like protocol, where the conflicts are resolved by using backoff mechanisms. We call this scheme Backoff-MAC (BO-MAC) and the main assumption is that each source node maintains a Congestion Window (CW) equal to 32, which doubles after collisions. However, BO-MAC achieves better performance comparing to the legacy IEEE 802.11 Standard since the Inter Frame Space (IFS) times are omitted, while the backoff mechanisms are used only in particular cases, i.e. when more than one node has the greatest impact in the network.

\subsection{Performance Results}
\begin{figure}[htb]
\centering
\includegraphics[width=1\columnwidth]{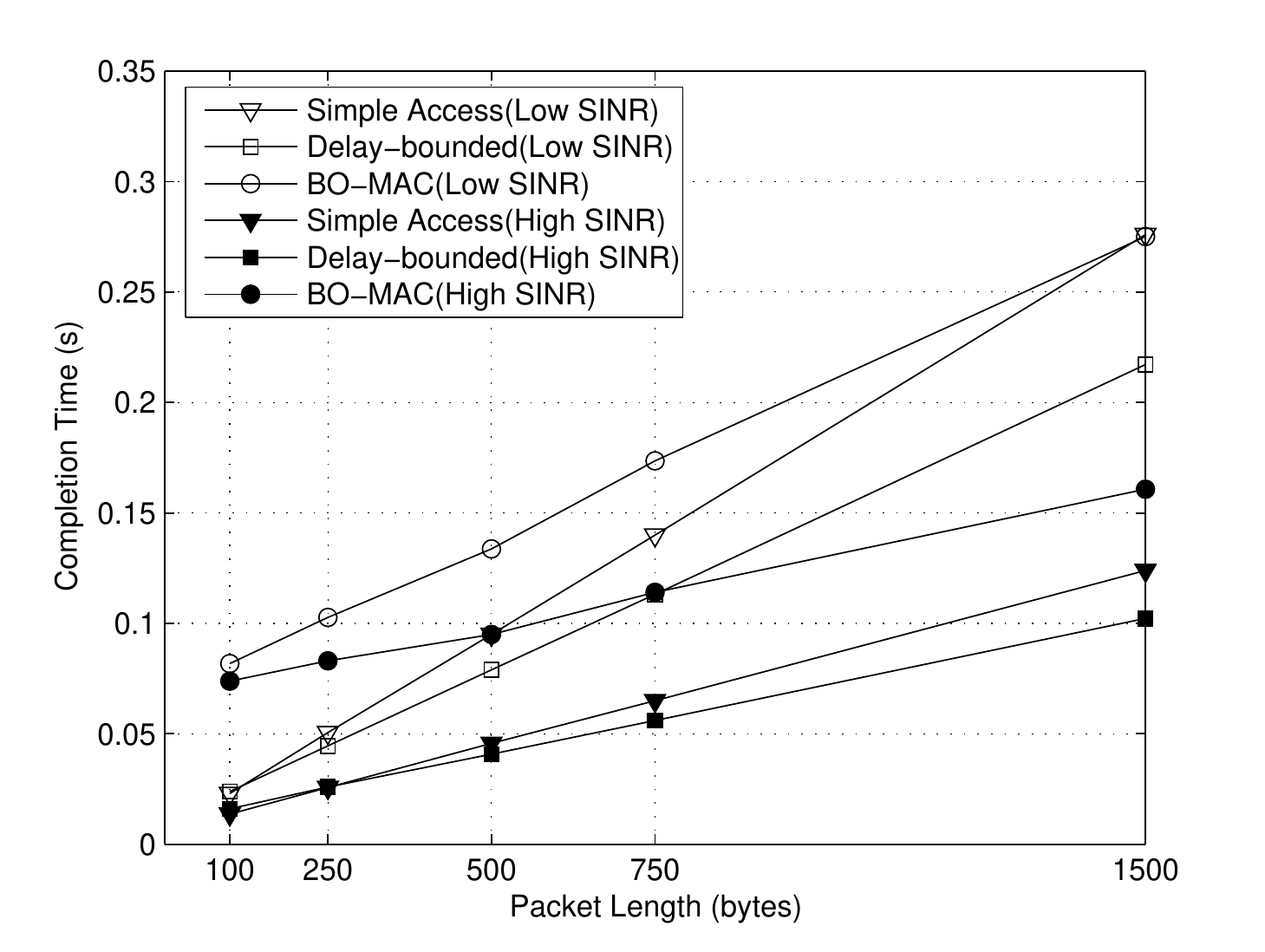}
\caption{Data Dissemination Completion Time (Game Theoretic Strategies vs. Backoff Strategy)}\label{f7}
\end{figure}

Figure \ref{f7} depicts the completion time of the data dissemination under both the proposed game theoretic schemes and the Backoff-MAC. The experiments have been conducted considering packets of various lengths (100-1500 bytes) for both \emph{low} and \emph{high SINR} conditions. First, we can see that our strategies outperform BO-MAC in all cases. However, it is worth noticing that game theoretic approaches achieve better performance even for worse SINR conditions for small data packets ($<$500 bytes). Furthermore, we observe that applying the delay-bounded strategy we are able to reduce the completion time up to 30\% (packet length of 1500 bytes in \emph{low SINR} conditions) compared to the simple access strategy. On the other hand, using high transmission data rates the gain is less significant, while for small packets ($<$250 bytes) the simple access scheme achieves better performance than the delay-bounded access strategy. This can be explained by the fact that the high transmission probabilities in the simple access scheme lead to an increased number of collisions. However, the small packet length implies small transmission time and, therefore, the collisions do not significantly affect the total completion time of the dissemination.

\begin{figure}[htb]
\centering
\includegraphics[width=1\columnwidth]{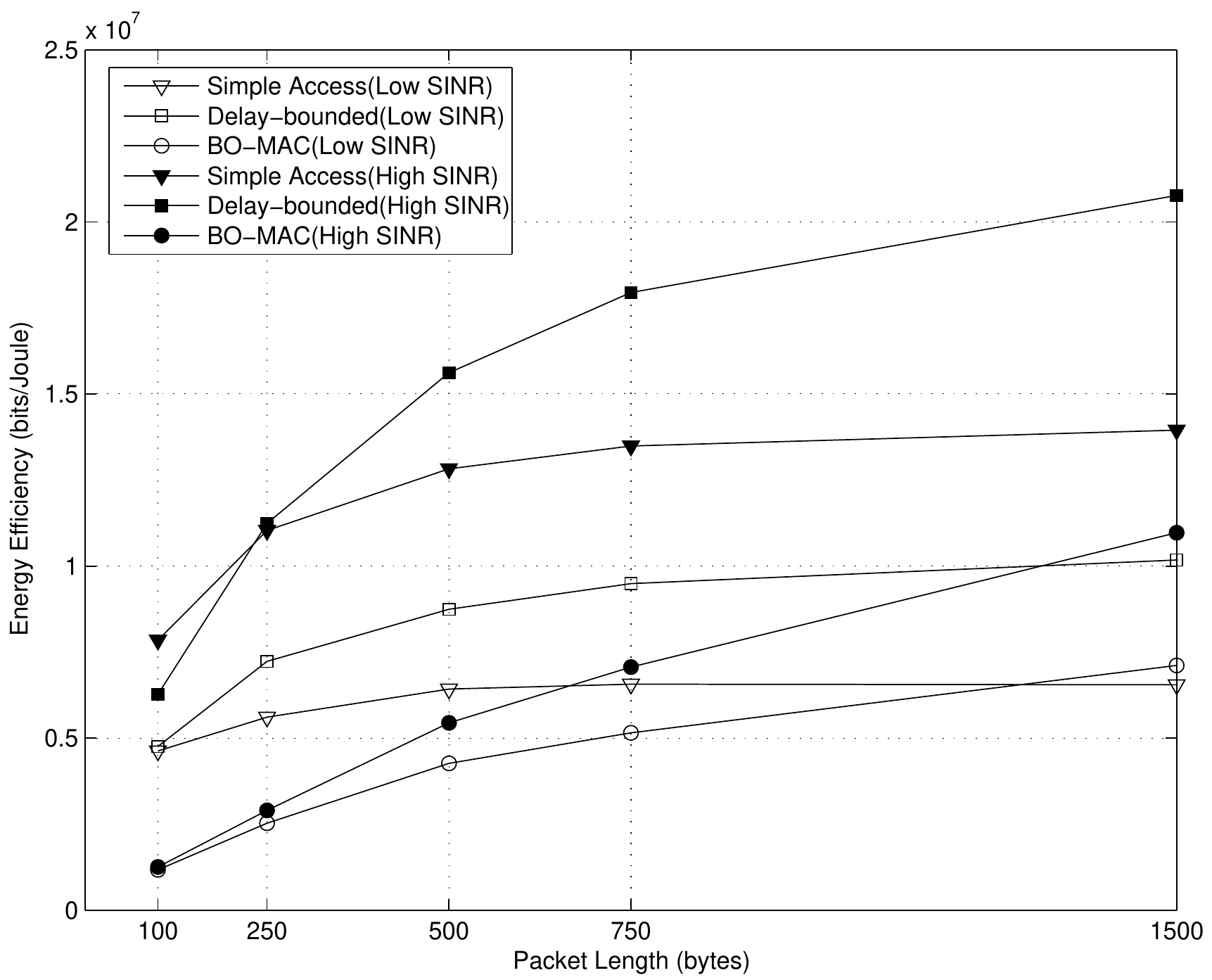}
\caption{Energy Efficiency (DAG Schemes vs. Backoff-MAC)}\label{f9}
\end{figure}

Figure \ref{f9} presents the simulation results with regard to the energy efficiency \cite{IEEEhowto:energy} of both versions of the Dissemination Access Game along with the Backoff-MAC. Several worthwhile observations are derived by this figure. First, although the completion time increases by using packets of big size, the energy efficient increases as well, since the amount of data delivered is beneficial for the network. Second, using the delay-bounded strategy in \emph{high SINR} conditions, we have a great enhancement up to 100\% compared to the BO-MAC scheme. In general, the two game theoretic schemes are proven to be more energy efficient than the optimized 802.11-like protocol. However, in the \emph{low SINR} scenario, BO-MAC slightly outperforms the simple access strategy for packets of 1500 bytes. This is the only case where it is more efficient to resolve the conflicts using congestion windows instead of applying game theoretic techniques. Nevertheless, in this case the delay-bounded strategy could provide an enhancement up to 50\% in terms of energy efficiency comparing to the other two methods.

\section{Concluding Remarks}
\label{sec:conclusion}
In this paper we proposed two game theoretic medium access strategies for data dissemination scenarios. Compared to an optimized IEEE 802.11-like scheme (BO-MAC), our approaches achieved up to 100\% enhancement in energy efficiency of the system, without degrading the offered Quality of Service. In our future work we are planning to elaborate on conflicts between more than two players in order to generalize our schemes, while we aim at the validation of our results by using analytical tools and methods.

\section*{Acknowledgments}

This work has been funded by the Research Projects CO2GREEN (TEC2010-20823), GREENET (264759) and Green-T (CP8-006).

\ifCLASSOPTIONcaptionsoff
  \newpage
\fi


\begin{thebibliography}{1}
\bibitem{IEEEhowto:nc1}
R. Ahlswede, Ning Cai, S.-Y.R. Li, R.W. Yeung, ``Network Information Flow",  IEEE Transactions on Information Theory, vol.46, no.4, pp.1204-1216, Jul 2000.
\bibitem{IEEEhowto:nc2}
S.-Y. R. Li, R. W. Yeung, Ning Cai, ``Linear network coding", IEEE Transactions on Information Theory, vol.49, no.2, pp.371-381, Feb. 2003.
\bibitem{IEEEhowto:coupon}
W. Feller, ``An Introduction to Probability Theory and Its Applications", volume 1. J. Wiley \& Sons, New York, 1964.
\bibitem {IEEEhowto:deb}
S. Deb, M. Medard, C. Choute, ``Algebraic gossip: a network coding approach to optimal multiple rumor mongering", IEEE Transactions on Information Theory, vol.52, no.6, pp. 2486- 2507, June 2006.
\bibitem {IEEEhowto:lucani}
D.E. Lucani,  F.H.P. Fitzek, M. Medard, M. Stojanovic, ``Network coding for data dissemination: it is not what you know, but what your neighbors don't know", 7th International Symposium on Modeling and Optimization in Mobile, Ad Hoc, and Wireless Networks (WiOPT), pp.1-8, 23-27 June 2009.
\bibitem {IEEEhowto:gintis}
H. Gintis, ``Game Theory Evolving", Princeton, NJ: Princeton University Press, 2000.
\bibitem {IEEEhowto:nash}
J. F. Nash, ``Non-cooperative games", Ann. Math., vol. 54, pp. 286–295, 1951.
\bibitem{IEEEhowto:ebert}
J.-P. Ebert, S. Aier, G. Kofahl, A. Becker, B. Burns and A. Wolisz, ``Measurement and Simulation of the Energy Consumption of a WLAN Interface", Technical University of Berlin, Telecommunication Networks Group, Tech. Rep. TKN-02-010, June 2002.
\bibitem {IEEEhowto:mackenzie}
A. MacKenzie and L. DaSilva, ``Game Theory for Wireless Engineers", Morgan \& Claypool Publishers, 2006.
\bibitem{IEEEhowto:80211g}
IEEE 802.11g WG, Part 11: Wireless LAN Medium Access Control (MAC) and Physical Layer (PHY) specifications - Amendment 4: Further Higher Data Rate Extension in the 2.4 GHz Band, June 2003.
\bibitem{IEEEhowto:energy}
M. Zorzi and R.R. Rao, ``Energy-constrained error control for wireless channels", IEEE Personal Communications, vol.4, no.6, pp.27-33, Dec 1997.

\end{thebibliography}
\end{document}